\documentclass[prl,aps,tightenlines,floatfix,showpacs,twocolumn]{revtex4}
\usepackage{graphics}
\usepackage{bm}
\usepackage{amsmath}
\usepackage{amssymb}

\begin{document}

\title{Coupled-channel evaluations of cross sections for scattering
       involving particle-unstable resonances}

  \author{P. Fraser$^{(1)}$}
  \email{pfraser@ph.unimelb.edu.au}  
  \author{K. Amos$^{(1)}$}
  \author{L. Canton$^{(2)}$}
  \author{G. Pisent$^{(2)}$}
  \author{S. Karataglidis$^{(3)}$}
  \author{J. P. Svenne$^{(4)}$}
  \author{D. van der Knijff$^{(5)}$}
  \affiliation{$^{(1)}$ School  of Physics,  University of  Melbourne, 
    Victoria 3010, Australia}
  \affiliation{$^{(2)}$ Istituto  Nazionale  di  Fisica  Nucleare,  
    Sezione  di Padova, \\  e Dipartimento di Fisica  dell'Universit\`a 
    di Padova, via Marzolo 8, Padova I-35131, Italia}
  \affiliation{$^{(3)}$ Department of Physics and Electronics, Rhodes 
    University, Grahamstown 6140, South Africa}
  \affiliation{$^{(4)}$ Department  of  Physics  and Astronomy,  
    University  of
    Manitoba,   and  Winnipeg   Institute  for   Theoretical  Physics,
    Winnipeg, Manitoba, Canada R3T 2N2}
  \affiliation{$^{(5)}$ Advanced  Research   Computing,  Information  
    Division, University of Melbourne, Victoria 3010, Australia}

\date{\today}

\begin{abstract}
How does the scattering cross section change when the colliding bound-state
fragments are allowed particle-emitting resonances?
This question is explored in the framework of a multi-channel algebraic 
scattering method of determining 
nucleon-nucleus cross sections at low  energies. Two
cases   are    examined,    the    first   being a \textit{gedanken}
investigation in which $n$+$^{12}$C scattering is  studied  with the 
target states assigned artificial widths.   The second is a study of
neutron   scattering  from $^{8}$Be;  a  nucleus  that  is  particle 
unstable.   Resonance character of the target states markedly varies
evaluated cross sections  from  those  obtained  assuming  stability
in the target spectrum.
\end{abstract}


\maketitle

The ready availability of radioactive ion beams\linebreak allows experimental
information to be obtained on many exotic nuclei, allowing for
study of novel structures, such as skins and halos. 
Of particular interest are the data obtained from scattering exotic nuclei from
hydrogen targets, which equates to proton scattering from those nuclei in the 
inverse kinematics. Such data have been analysed in terms of effective 
nucleon-nucleus interactions used in distorted wave approximations 
\cite{Am00,La01,St02}, or using coupled-channels approaches \cite{Ri98,Am03}.

This paper considers the situation of low-energy neutron scattering from
two light mass nuclei ($^{12}$C and $^8$Be) for which discrete resonance 
effects in the elastic
cross section are usually present. Such resonance properties most often
result  from  channel  coupling  and  are  reproduced here using  
a multi-channel  algebraic  scattering  (MCAS) theory~\cite{Am03}.
With MCAS, solutions of coupled Lippmann-Schwinger equations are found (in 
momentum
space) by using finite-rank separable representations  of  an  input 
matrix of nucleon-nucleus interactions. An ``optimal'' set of sturmian 
functions~\cite{We65} is used as the expansion set.  Details are given
in Refs.~\cite{Am03,Ca05}. The advantages of using the MCAS 
method include an ability to locate all compound system resonance centroids and
widths regardless of how narrow those resonances 
may be, and, by use of orthogonalizing pseudo-potentials (OPP) in 
generating sturmians, to ensure the Pauli principle is not 
violated~\cite{Ca05}, despite the collective model formulation of 
nucleon-nucleus interactions used therein. The latter is of paramount 
importance for coupled-channel
calculations~\cite{Am05}, as otherwise some compound nucleus states 
so defined possess spurious components in their wave functions.

MCAS is used to find solutions of the coupled-channel, partial-wave 
expanded Lippmann-Schwinger equations for each total system spin-parity
($J^\pi$),
\begin{multline}
T^{J^{\pi}}_{cc'}( p, q; E ) = 
  V^{J^{\pi}}_{cc'}( p, q )  \\ 
+ \mu  \left[ \sum^{\text{open}}_{c'' = 1} \int^{\infty}_0
V^{J^{\pi}}_{cc''}( p, x) \frac{ x^2 }{ k^2_{c''} - x^2 + i\varepsilon
  } T^{J^{\pi}}_{c''c'}( x, q; E ) \, dx \right. 
 \\
 - \left. \sum^{\text{closed}}_{c'' = 1} \int^{\infty}_0
  V^{J^{\pi}}_{cc''}( p, x ) \frac{ x^2 }{ h^2_{c''} + x^2 }
  T^{J^{\pi}}_{c''c'}( x,q; E ) \, dx \right] \, ,
\label{LScc'}
\end{multline}
where a finite set of scattering channels,  denoted  $c$, are considered,
and where $\mu =\textstyle\frac{2\overline{m}}{\hbar}$, ${\overline m}$ being 
the reduced mass. There are two summations as the open and closed 
channel components are separated, with wave numbers
\begin{equation}
k_c = \sqrt{\mu(E - \epsilon_c)} \: \text{\ and\ } \: h_c = 
\sqrt{\mu(\epsilon_c - E)}\; ,
\label{kandh}
\end{equation}
for $E > \epsilon_c$ and $E < \epsilon_c$ respectively. $\epsilon_c$
is the energy threshold at which channel $c$ opens (the excitation 
energies of the target nucleus).  Henceforth the $J^\pi$ superscript 
is to be understood. Expansion of $V_{cc'}$ in terms of a finite 
number ($N$) of sturmians leads to a separable representation of the 
scattering matrix~\cite{Am03}
\begin{multline}
S_{cc'} =\; \delta_{cc'} \; -\, i^{(l_{c'} - l_c + 1)}\; \pi\; \mu 
\\
\times 
\sum^{N}_{n,n'=1} 
\sqrt{k_c} \hat{\chi}_{cn}(k_c) \left( \left[ {\mbox{\boldmath
$\eta$}} - \mathbf{G}_0 \right]^{-1} \right)_{nn'} \hat{\chi}_{c'n'}
  \sqrt{k_{c'}}\; ,
  \label{Smatrix}
\end{multline}
where $c$ and $c'$ refer now only to open channels, $l_c$ is the 
partial wave with channel $c$  and  the  Green's  function matrix is
\begin{multline}
  \left[ \mathbf{G}_0 \right]_{nn'} = \mu \left[
  \sum^{\text{open}}_{c = 1} \int^{\infty}_0 \hat{\chi}_{cn}(x)
  \frac{x^2}{ k^2_c - x^2 } \hat{\chi}_{cn'}(x) \, dx
  \right. \\
   - \left. \sum^{\text{closed}}_{c = 1}
  \int^{\infty}_{0} \hat{\chi}_{cn}(x) \frac{ x^2 }{ h^2_c + x^2 }
  \hat{\chi}_{cn'}(x) \, dx \right].
  \label{S-Gmod}
\end{multline}
\mbox{\boldmath $\eta$}  is a column vector of sturmian eigenvalues 
and $\hat \chi$ are form factors determined from the chosen sturmian
functions. Details are given in Ref~\cite{Am03}.

Traditionally, all target states are taken to have eigenvalues of zero width 
and the (complex)  Green's  functions are  evaluated using the method of 
principal parts. This assumes time evolution of target states is 
given by
\begin{equation}
\left|x,t\right\rangle = e^{-i H_0 t/\hbar}\; \left|x,t_0\right\rangle = 
e^{-i E_0 t/\hbar}\; \left|x, t_0 \right\rangle.
\end{equation}
However, if states decay, they evolve as~\cite{Wo90}
\begin{equation}
\left|x,t\right\rangle = e^{-\textstyle\frac{\Gamma}{2}t} \; e^{-iE_0t/\hbar}\; 
\left|x,t_0\right\rangle.
\end{equation}

Thus, in the Green's function, channel energies become complex, 
as do the squared channel wave numbers,
\begin{equation}
\hat{k_c}^2 = \mu \left(E - \epsilon_c + \textstyle\frac{i\Gamma_c}{2} 
\right) \: ; \:\:
\hat{h_c}^2 = \mu \left( \epsilon_c - E - \textstyle\frac{i\Gamma_c}{2}
\right) ,
\label{hatkandhath}  
\end{equation}  
where $\textstyle\frac{ \Gamma_c }{2}$ is half the width of the target  state 
associated with channel $c$. Thus, the Green's function matrix 
elements are
\begin{multline}
  \left[ \mathbf{G}_0 \right]_{nn'}  = \mu \left[
  \sum^{\text{open}}_{c = 1} \int^{\infty}_0 \hat{\chi}_{cn}(x)
  \frac{x^2}{ k_c^2 - x^2 + \textstyle\frac{ i \mu \Gamma_c }{2} } 
  \hat{\chi}_{cn'}(x) \, dx
  \right. \\
  - \left. \sum^{\text{closed}}_{c = 1}
  \int^{\infty}_{0} \hat{\chi}_{cn}(x) 
  \frac{ x^2 }{ h_c^2 + x^2 - \textstyle\frac{ i \mu \Gamma_c }{2} }
  \hat{\chi}_{cn'}(x) \, dx \right] ,
  \label{S-Ggamma2}
\end{multline}
where $k_c$ and $h_c$ are as in Eq.~(\ref{kandh}).
Thus, poles are moved significantly off the real axis, and integration
of  a  complex  integrand  along the real momentum axis is feasible.
This has been done; however, for any infinitesimal-width target state, or 
resonance so narrow that it can be treated as such, the method of
principal parts has been retained.

\begin{figure}[h]
\begin{center}
\scalebox{0.37}{\includegraphics*{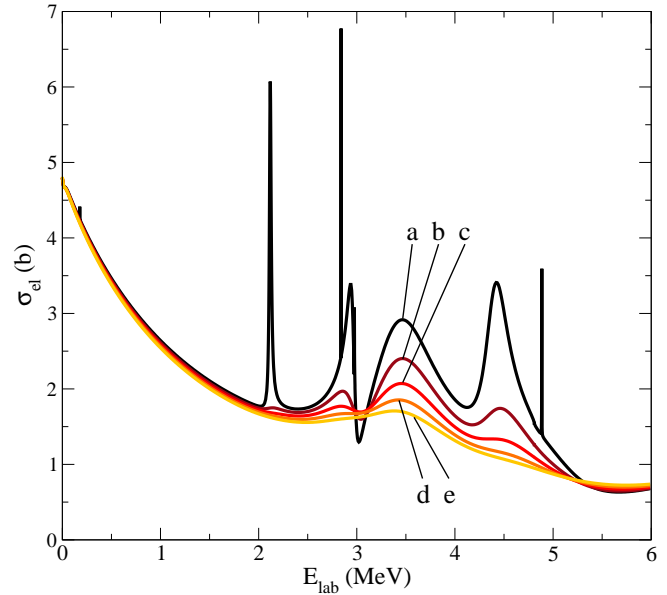}}
\end{center}
\caption{ \label{C12+n}
(Color online) Calculated cross sections for hypothetically $n$-(unstable) 
$^{12}$C scattering as functions of neutron energy. Labels as per 
Table~\ref{C12WidthTable}.}
\end{figure}

As previously~\cite{Am03,Ca05}, the $^{13}$C ($n + ^{12}$C) 
system is studied using the  MCAS  approach  with a rotational 
model prescription of the matrix  of interaction potentials 
connecting three states of $^{12}$C (the $0^+_{g.s.},\; 2^+_1$ (4.43~MeV) and 
$0^+_2$ (7.64~MeV)), using the same interaction Hamiltonian and allowing for 
Pauli blocking via the OPP scheme. In the first instance, all three states are
considered zero-width, giving the elastic scattering 
cross section of neutrons to 6~MeV as previously published.
Additionally, evaluations are made for the same interaction
allowing the $2^+_1$ and $0_2^+$ states of $^{12}$C to 
have particle emission widths of varying size;  the ground state kept with zero
width. Results are displayed in Fig.~\ref{C12+n}.

Cross sections follow a marked trend as hypothetical state widths increase.
Widths used are listed in Table~\ref{C12WidthTable}.
\begin{table}[h]
\begin{ruledtabular}
\caption{
\label{C12WidthTable}
Artificial widths ($\Gamma$, in MeV) assigned to $^{12}$C eigenstates.} 
\begin{tabular}{cccc}
Curve   & $0^+_1$ width & $2^+_1$ width & $0^+_2$ width\\
\hline
a   & ---       & ---       & --- \\
b   & 0.00      & 0.20      & 0.60\\
c   & 0.00      & 0.40      & 1.20\\
d   & 0.00      & 0.60      & 1.80\\
e   & 0.00      & 0.80      & 2.40\\
\end{tabular}
\end{ruledtabular}
\end{table}
Ascribing the first widths to the excited states (set (b)), very 
narrow resonances in the original cross section disappear.  From the earlier
studies~\cite{Am03,Ca05}, it is noted that those (narrow) compound 
resonance states are dominated  by the  coupling  of  an  $sd$-shell
nucleon to the $2^+_1$ state in $^{12}$C. The broader resonances 
remain evident in the cross section as the state 
widths are artificially increased. However, with these increases the 
remaining resonances smear out. In the case of the broadest target 
states (set (e)) the cross section has very little  remnant  of  the  compound 
system resonances. Clearly only the cross section from evaluation 
with three zero-width target states replicates measurement.

Table~\ref{C13-spect} displays the widths of states in the compound nucleus,
$^{13}$C, found using MCAS when attributing the diverse widths to the 
excited states of $^{12}$C listed in Table~\ref{C12WidthTable}. The
first column after $J^{\pi}$ lists the bound state and resonance centroid 
energies obtained from the calculation made with the physically  
reasonable, zero-width excitation energies of the $2^+_1$ and 
$0^+_2$ states of $^{12}$C. Allowing those states to be resonances with the 
widths selected alters the state energy centroids by at most a few tens of 
keV, so these are not listed. 
\begin{table}[h]
\begin{ruledtabular}
\caption{\label{C13-spect} Widths (in MeV) for $^{13}$C states from
calculation allowing states of $^{12}$C to be resonances
as listed in Table~\ref{C12WidthTable}.}
\begin{tabular}{ccccccc}
$J^\pi$ & Centroid & (a) & (b) & (c) & (d) & (e)\\
\hline
$\frac{1}{2}^-$ & $-$4.82  & --- & --- & --- & --- & --- \\
$\frac{1}{2}^+$ & $-$2.04  & --- & --- & --- & --- & --- \\
$\frac{5}{2}^+$ & $-$1.85  & --- & --- & --- & --- & --- \\
$\frac{3}{2}^-$ & $-$1.36  & --- & --- & --- & --- & --- \\
$\frac{5}{2}^-$ & \ \ 0.16 &$7\times 10^{-10}$ & 0.09 & 0.18 & 0.28 & 0.37 \\
$\frac{5}{2}^+$ & \ \ 1.95 &\ \ 0.01      & 0.10 & 0.19 & 0.28 & 0.38 \\
$\frac{7}{2}^+$ & \ \ 2.62 &$9\times 10^{-7}$ & 0.09 & 0.18 & 0.28 & 0.37 \\
$\frac{3}{2}^+$ & \ \ 2.74 &\ \ 0.04      & 0.13 & 0.23 & 0.33 & 0.42 \\
$\frac{1}{2}^-$ & \ \ 2.75 &$8\times 10^{-4}$ & 0.27 & 0.55 & 0.82 & 1.11 \\
$\frac{3}{2}^+$ & \ \ 3.25 &\ \ 0.45      & 0.50 & 0.56 & 0.61 & 0.67 \\
$\frac{5}{2}^+$ & \ \ 4.06 &\ \ 0.13      & 0.25 & 0.38 & 0.52 & 0.66 \\
$\frac{9}{2}^+$ & \ \ 4.51 &$7\times 10^{-4}$ & 0.09 & 0.19 & 0.28 & 0.38 \\
$\frac{1}{2}^+$ & \ \ 4.76 &\ \  0.52      & 0.71 & 0.92 & 1.14 & 1.37 \\
\end{tabular}
\end{ruledtabular}
\end{table}
Thus, it is observed that allowing these target states to be resonances 
mostly affects widths of the resulting compound nucleus resonances.  Those 
variations
are consistent with changes noted in the cross section, with
sharp resonances found for the zero-width state case rapidly disappearing
and the others broadening to an extent that only a few are left 
distinguishable from a background. It is important to note, though, that
all states in the compound system defined by the 
coupled-channel evaluations remain present, with, in this case, 
centroid energies little affected but widths increased. 

\begin{figure}[h]
\begin{center}
\scalebox{0.54}{\includegraphics*{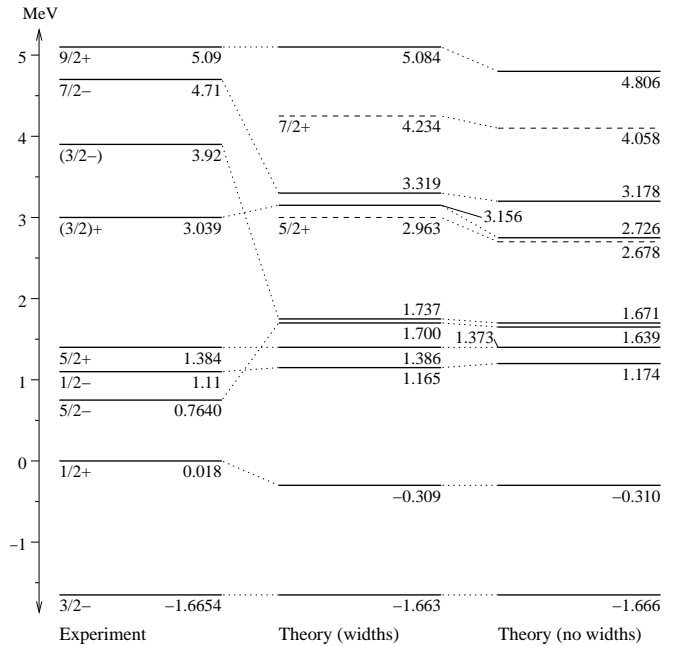}}
\end{center}
\caption{ \label{Be8+n-spec}
Experimental $^9$Be spectrum and that calculated from neutron 
scattering with stable and unstable $^8$Be.}
\end{figure}

The low excitation $^8$Be spectrum has a $0^+$ ground state that 
has a small width for its decay into two $\alpha$-particles 
($6 \times 10^{-6}$ MeV), a broad $2^+$ resonance state with centroid at 3.03
MeV and  
width of 1.5 MeV, followed by a broader  $4^+$ resonance state 
with centroid at 11.35 MeV and width ${\sim}3.5$ MeV~\cite{Ti04}.
Two evaluations of the $n + ^8$Be cross section are obtained with MCAS; in 
both, the ground state is taken as having zero-width. In 
the  first, both the $2^+$ and $4^+$ states are also taken as zero-width   
(ignoring their known $\alpha$-decay
widths) whereas in the second, the empirical widths are used.
In both calculations, the same nuclear interaction is considered. It
is taken from a rotor model with parameter values chosen in the 
finite-width states calculation to reproduce some aspects of the 
experimentally determined structure of $^9$Be~\cite{Ti04}, shown graphically
in Fig.~\ref{Be8+n-spec}. 

\begin{figure}[h]
\begin{center}
\scalebox{0.37}{\includegraphics*{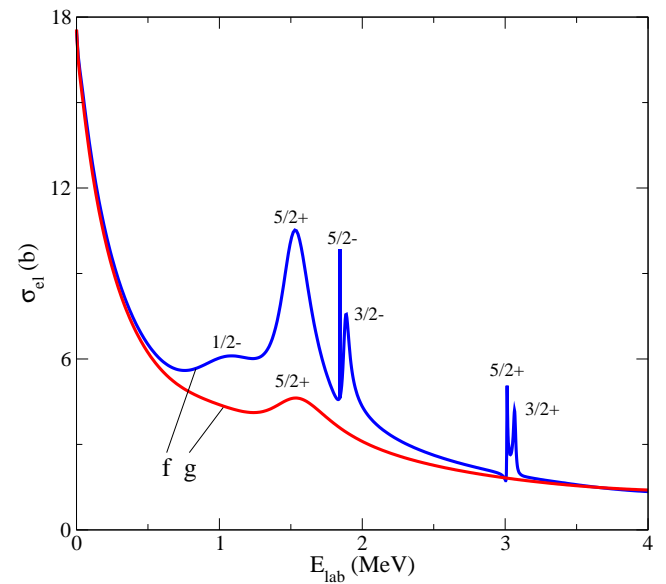}}
\end{center}
\caption{ \label{Be8+n}
(Color online) Calculated cross sections for neutron scattering from (f) 
stable and (g) unstable $^8$Be as functions of neutron energy.}
\end{figure}

The results for the scattering cross sections are shown
in Fig.~\ref{Be8+n}. Upon introducing target state widths, 
as found in the $n + ^{12}$C investigation, the resonances are 
suppressed but still present; their widths increasing and magnitudes 
decreasing so as all but the $\textstyle\frac{5}{2}^+$ cannot be discerned 
from the
background. Compound system resonances of both calculations are at essentially 
the same energies.  These effects are further illustrated by the centroid 
energies and widths of the resonances listed in Table~\ref{Be9-spect}.
\begin{table}[h]
\begin{ruledtabular}
\caption{\label{Be9-spect} $^9$Be state centroids and widths ($E$ \& $\Gamma$ in
MeV) from calculation with $^8$Be states taken as zero-width
and then with known resonance widths, and experimental widths.}
\begin{tabular}{cccccc}
& \multicolumn{2}{c}{Zero-width} & \multicolumn{2}{c}{Resonances} &Experiment\\
$J^\pi$ & $E$(f) & $\Gamma$(f) & $E$(g) & $\Gamma$(g) & $\Gamma$(exp.) \\
\hline
$\frac{3}{2}^-$ & $-$1.67 & ---   & $-$1.66 & ---   & ---  \\
$\frac{1}{2}^+$ & $-$0.31 & ---   & $-$0.31 & ---   & 0.217$\pm$0.001\\
$\frac{1}{2}^-$ & \ \ 1.17 & 0.646 & \ \ 1.16 & 0.972 & 1.080$\pm$0.110\\
$\frac{5}{2}^+$ & \ \ 1.37 & 0.118 & \ \ 1.39 & 0.244 & 0.282$\pm$0.011\\
$\frac{5}{2}^-$ & \ \ 1.64 & 3.7$\times10^{-9}$ & \ \ 1.70 & 0.694 & $7.8{\times}10^{-4}$\\
$\frac{3}{2}^-$ & \ \ 1.67 & 0.022 & \ \ 1.74 & 0.682 & 1.330$\pm$0.360\\
$\frac{5}{2}^+$ & \ \ 2.68 & 0.003 & \ \ 2.96 & 1.141 & N/A\\
$\frac{3}{2}^+$ & \ \ 2.73 & 0.009 & \ \ 3.16 & 1.856 & 0.743$\pm$0.055\\
$\frac{7}{2}^-$ & \ \ 3.18 & 0.009 & \ \ 3.32 & 0.786 & 1.210$\pm$0.230\\
$\frac{7}{2}^+$ & \ \ 4.06 & 0.072 & \ \ 4.23 & 0.873 & N/A\\
$\frac{9}{2}^+$ & \ \ 4.81 & 0.189 & \ \ 5.08 & 1.261 & 1.330$\pm$0.090\\
\end{tabular}
\end{ruledtabular}
\end{table}
Column 1 (after $J^{\pi}$) lists the resultant centroid 
energies of the spectrum found assuming the $2_1^+$ and $4^+_1$ excited states 
in $^8$Be have zero width, and the widths of these resonances
(energies above the $n + ^8$Be threshold) are in column 2.
Columns 3 and 4 list energies and widths, respectively,
obtained using the MCAS scheme on allowing the two excited states
in $^8$Be to have their ascribed widths. Column 5 lists the experimental 
widths~\cite{Ti04}.
As found in the study of the $^{13}$C spectrum, taking the excited states
of $^8$Be to be resonances gives the same spectral list as when they
are treated as zero-width, but the widths of the compound nuclear states found
significantly increase. This is again reflected in the 
cross sections. These increases bring the theoretical $^9$Be state widths 
closer, often significantly, to experimental values. In this case, some 
centroid energies are shifted by up to 330 keV. 

In conclusion, a multi-channel algebraic scattering approach to theoretical 
evaluation nucleon-nucleus scattering information has been extended to consider
widths of target nucleus eigenstates. Resultant resonances in obtained cross 
sections suffer only minor changes to their energy centroids.
However, their widths are substantially increased, often making them 
indistinguishable from the scattering background, but closer to experimentally
determined widths. This was observed for $n + ^8$Be scattering using 
calculations with and without experimental target state widths. Furthermore, it
is found that this effect increases as target state widths increase, with sharp
resonances quickly obscured. This was observed from calculations using a
series of artificial target state widths.  

\begin{flushleft}
\section*{Acknowledgements}
\end{flushleft}

This research was supported in part by Melbourne University Scholarship Office 
PORES program. P.F. acknowledges the gracious hospitality of the Department of 
Physics and Astronomy, University of Manitoba; the INFN, Sezione di Padova, 
Padova and the Dipartimento di Fisica dell' Universit\'a di Padova; and the 
Department of 
Physics and Electronics, Rhodes University. S.K. acknowledges support 
from the National Research Foundation (South Africa). J.P.S. acknowledges 
support
from the Natural Sciences and Engineering Research Council (NSERC), Canada.

\bibliography{Fraser-PRL}

\end{document}